\documentstyle[12pt]{article}
\begin{document}
\vspace{17mm}
\normalsize
\begin{center}
{\Large \bf POLYNOMIAL LIE ALGEBRAS $sl_{pd}(2)$ IN ACTION:
SMOOTH $sl(2)$ MAPPINGS AND APPROXIMATIONS}\\
\vspace{8mm}
               V.P. KARASSIOV\\
{\it Lebedev Physical Institute, Leninsky prospect 53, 117924 Moscow, Russia\\
Internet: karas@sci.fian.msk.su}
\end {center}
\begin{abstract}
We examine applications of polynomial Lie algebras $sl_{pd}(2)$
to solve physical tasks in $G_{inv}$-invariant models of coupled
subsystems in quantum physics. A general operator formalism is
given to solve spectral problems using expansions of generalized coherent
states, eigenfunctions and other physically important quantities by power
series in the $sl_{pd}(2)$ coset generators $V_{\pm}$. We also discuss some
mappings and approximations related to the familiar $sl(2)$ algebra formalism.
On this way a new closed analytical expression is found for energy spectra
which coincides with exact solutions in certain cases and, in general,
manifests an availability of incommensurable eigenfrequencies related to
a nearly chaotic dynamics of systems under study.
\end{abstract}

\section{Introduction. General remarks}

Recently, in a series of our papers /1-4/ a new efficient Lie-algebraic
approach has been suggested to solve both spectral and evolution problems
for some nonlinear $G_{inv}$-invariant models of coupled subsystems in
quantum physics. It was based on exploiting a formalism of polynomial
Lie algebras $g_{pd}$ as dynamic symmetry algebras $g^{DS}$ of models
under study, and besides generators of these algebras $g_{pd}$ can be
interpreted as $G_{inv}$-invariant "essential collective dynamic variables"
in whose terms model dynamics are described completely. Specifically,
this approach enabled to develop some efficient techniques for solving
physical tasks in the case $g^{DS}=sl_{pd}(2)$ when model Hamiltonians $H$
are expressed as follows
%%%%%%%%%%%%%%
$$ H = aV_0 +g V_+ + g^* V_- +C= {\bf A V}+C ,\quad [V_{\alpha}, C]=0,
\quad V_- =(V_+)^+,      \eqno (1.1) $$
%%%%%
where $C$ is a function of $G_{inv}$-dependent model integrals of
motion $R_i$ and $V_0, V_{\pm}$ are the $sl_{pd}(2)$ generators satisfying
the commutation relations
%%%%%%%%%
$$ [V_0, V_{\pm}]= \pm V_{\pm}, \quad [V_-, V_+] =  \psi_n(V_0+1) -
\psi_n(V_0),   \eqno (1.2)$$
%%%%%%%%
with the structure function $\psi_n(V_0)$ being a polynomial $\psi_n(V_0)=
A(\{R_j\})\prod_{i=1}^{n} (V_0- \lambda_i(\{R_j\}))$ of the degree $n$ in
$V_0$.

For example, the three-boson model Hamiltonian
%%%%%%%%%
$$ H_2 = \omega_1 a^+_1 a_1 +\omega_2 a^+_2 a_2 +\omega_3 a^+_3 a_3 +
g(a^+_1a^+_2) a_3 + g^*(a_1a_2) a_3 ^+      \eqno (1.3)$$
%%%%%%%
can be expressed in the form (1.1) if using the substitutions
%%%
$$V_0 =(N_1+N_2-N_3)/3,\; V_+ =(a^+_1a^+_2) a_3, \quad a= \omega_1 +\omega_2-
\omega_3, \quad N_i=a^+_i a_i, $$
$$2C =R_1(\omega_1 -\omega_2) +R_2(\omega_1+\omega_2 +2\omega_3),
\; R_1= N_1-N_2,\;3R_2= N_1+N_2+2N_3 \eqno (1.4a)$$
%%%%%%%
In this case the structure function $\psi_n(V_0)\equiv\psi_3(V_0)$ is
given as follows
%%%%
$$ \psi_3(V_0)=\frac{1}{4} (2V_0 +R_2-R_1)(2V_0 +R_1+R_2)(-V_0 +R_2+1)
\eqno (1.4b)$$
%%%%%%%

All techniques developed were based on using expansions of physically
important quantities (evolution operators, generalized coherent states (GCS),
eigenfunctions etc.) by power series in the $sl_{pd}(2)$ coset generators
$V_{\pm}$. Besides, in the Schroedinger picture one has exploited
decompositions
%%%%%%%%
$$ L(H) =\sum_{\oplus}L([l_i]), \quad (V_+V_- -\psi_n(V_0))|_{L([l_i])}=0
 \eqno (1.5)$$
%%%%%
of Hilbert spaces $L(H)$ of quantum states of model in direct sums of
the subspaces $L([l_i])$ which are irreducible with respect to joint actions
of algebras $sl_{pd}(2)$ and groups $G_{inv}$ and describe specific
"$sl_{pd}(2)$-domains" evolving independently in time under  action of
the Hamiltonians (1.1). The subspaces $L([l_i])$ are spanned by basis vectors
%%%%%%%%%%%%%%%%
$$|[l_i];v\rangle=[(\psi_n(l_0+v))^{(v)}]^{-1/2}V_+^v|[l_i]\rangle,
\quad (\psi_n(x))^{(v)}\equiv \prod_{r=0}^{v-1}\psi_n (x-r),$$
$$V_0|[l_i];v\rangle=(l_0+v)|[l_i];v\rangle, \; R_i|[l_i];v\rangle
=l_i|[l_i];v\rangle, \; \psi_n(R_0)\equiv \psi_n (V_0)-V_+V_-
\eqno (1.6)$$
%%%%%%%%%
where $|[l_i]\rangle$ is the lowest vector ($ V_-|[l_i]\rangle=0,
\psi_n(l_0)=0$) of $L([l_i])$.

Then, using Eqs. (1.1)-(1.2) one may get Jacobi-type three-term recurrence
relations for amplitudes $Q_v(E_f) \equiv\langle [l_i]; v|E_f\rangle$ of
expansions of energy eigenstates $|E_f\rangle$ in bases $\{|[l_i]; v
\rangle\}$. Besides, energy spectra $\{E_f\}$ of bound states are given by
roots of certain spectral functions (polynomials for the compact version
$sl_{pd}(2)=su_{pd}(2)$) which are determined for given structure
functions $\psi_n(x)$ with the help of similar recurrence relations /1-3/.
Another way, exploiting the Bargmann-type representation of the $sl_{pd}(2)$
generators,
%%%%
$$ V_+=z,\; V_0=zd/dz+l_0,\; V_-= z^{-1}\psi_n(zd/dz+l_0),  \eqno (1.7a)$$
%%%%
reduces these tasks to solving some singular differential equations /1-3/.
When using a conjugate to (1.7a) representation of the $sl_{pd}(2)$
generators,
%%%%%
$$ V_-=d/dz,\; V_0=zd/dz+l_0,\; V_+=\psi_n(zd/dz+l_0)(d/dz)^{-1}
\eqno (1.7b)$$
%%%
this way leads to solving the Riccati-type equations for structure functions
$\psi_3(x)$ of the degree $n=3$ (that is the case for the Hamiltonian (1.3)).
In the
 paper /4/
some integral expressions were found for amplitudes $Q_v(E)$, eigenenergies
$\{E_a\}$ and evolution operators
$U_{H}(t)$ with the help of a specific "dressing" (mapping) of solutions of
some auxiliary exactly solvable tasks with the dynamic algebra $sl(2)$.

However, all these and other results do not yield simple working formulas
for analysis of models (1.1) and revealing different physical effects (e.g.,
collapses and revivals of the Rabi oscillations /5/) at arbitrary initial
quantum states of models. Besides, solutions /4/
of spectral tasks manifest so-called "quantum discontinuities" /6/: a
disappearence of wave functions when attaining the limit of auxiliary
$sl(2)$ Hamiltonians that makes difficult to compare completely quantum
models with their semi-classical analogs. Therefore, it is necessary to
develop some simple techniques enabling to display important physical
peculiriaties of models (1.1)-(1.2). In the case $g^{DS}=sl(2)$, when the
structure functions $\psi_n(x)\equiv\psi_2(x)$ are quadratic functions
$\psi_2(x)=(j\pm x)(\mp j+1-x),\; l_0=\mp j$), the GCS formalism of the group
orbit type /7/ is known to be an efficient tool for analyzing both linear /7/
and non-linear /7-9/ models. This formalism based on properties of the
$SL(2)$ group displacement operators $S_V(\xi)=\exp(\xi V_+-\xi^* V_-),\;\xi
=r\exp(-i\theta)$ yields exact solutions /7/ for linear models and
variational schemes (corresponding to the Ehrenfest theorem) to obtain
effective mean-field approximate solutions for non-linear models /8-9/.

Below we examine some possibilities of generalizations of this formalism
for solving spectral problems of models (1.1)-(1.2) (Section 2) and give a
variational scheme to find "smooth" $sl(2)$-approximations of these solutions
(Section 3) using an isomorphism of the $sl_{pd}(2)$ algebras to special
subalgebras of the extended enveloping algebra ${\cal U}(sl(2))$ of the
familiar algebra $sl(2)$. This isomorphism is established via a generalized
Holstein-Primakoff mapping given as follows /1-3/
%%%%
$$Y_0 = V_0-l_0\mp j,\; j=\frac{s}{2}, \; Y_+= V_+
\sqrt{\frac{(j\mp Y_0)(\pm j+1+Y_0)}{\psi_n(V_0+1)}},\; Y_-=(Y_+)^+,
 \eqno (1.8)$$
%%%%%
where $ Y_{\alpha}$ are the $sl(2)$ generators, upper and lower signs
correspond to the $su(2)$ and $su(1,1)$ algebras respectively. In Section
4 some prospects of further studies along these lines are briefly outlined.

\section{A general operator formalism to solve spectral problems}

As is known /7/, the Hamiltonians (1.1) are simply diagonalized with the help
of operators
%%%%
$$ S_V(\xi)=\exp(\xi V_+-\xi^* V_-)=\exp[t(r)e^{i \theta} V_+]
\exp[-2\ln c(r) V_0]\exp[-t(r)e^{-i\theta} V_-],\quad \xi=re^{i \theta}
\eqno (2.1)$$
%%%%%%
when $V_{\alpha}$ are generators of the familiar $sl(2)$ algebra ($t(r)=
\tan r, c(r)=\cos r$ for $su(2)$ and $t(r)=\tanh r, c(r)=\cosh r$ for
$su(1,1)$). Indeed, using the well-known $sl(2)$ transformation properties
of operators $V_{\alpha}$ one finds the transformation
%%%%%%%%%%%
$$ H\longrightarrow \tilde{H}(\xi)=S_V(\xi)H S_V(\xi)^{\dagger}= C +
V_0 A_0(a,g;\xi)+V_+ A_+(a,g;\xi) +V_- A^*_+(a,g;\xi)  \eqno (2.2a)$$
%%%%%
of the Hamiltonians (1.1) under the action of operators $S_V(\xi)$. Then,
supposing $A_+(a,g;\xi)=0$ we find a value $\xi_0$ of the parameter
$\xi$ diagonalizing the Hamiltonian $\tilde{H}(\xi)$. For example,
in the case of the $su(2)$ algebra we have /7/
%%%%%%%%%%%%%
$$\tilde{H}(\xi_0)=S(\xi_0)H S(\xi_0)^{\dagger}= C + V_0\sqrt{a^2+4 |g|^2},
\quad \xi_0=\frac{g}{2|g|}arctg\frac{2|g|}{a}, \eqno (2.2¡)$$
%%%%%%%
and the corresponding eigenenergies $E([l_i];v;\xi_0)$ and eigenfunctions
$|[l_i];v;\xi_0 \rangle$ are expressed
as follows
%%%%%
$$a)\;E([l_i];v;\xi_0)= C + (-j+v)\sqrt{a^2+4 |g|^2},  \eqno (2.3a)$$
$$b)\;|[l_i];v;\xi_0\rangle= S_V(\xi_0)^{\dagger}|[l_i];v; \rangle=
\exp(-\xi_0 V_++ \xi_0^*V_-)|[l_i];v; \rangle=(\cos^2 r)^{j-v} \times$$
$$ \sum_{f\geq 0} \frac{(-e^{i \theta}tg r)^{f-v}}{(f-v)!}
{F(-v,-v+2j+1;f-v+1; \sin^2 r)} [\frac{(2j-v)!f!}{(2j-f)!v!}]^{1/2}
|[l_i];f;\rangle    \eqno(2.3b)$$
%%%%%%%%%%
where $F(...)$ is the Gauss hypergeometric function. An equivalent way /8/
to obtain the results (2.3) is based on using the stationarity conditions
$$\frac{\partial E([l_i];v;\xi)}{\partial \theta}=0,\quad
\frac{\partial E([l_i];v;\xi)}{\partial r}=0
\eqno (2.4)$$
%%%%%%%
for the energy functional $E([l_i];v;\xi)=\langle[l_i];v;\xi|H|[l_i];v;\xi
\rangle$ defined with the help of the $SL(2)$ GCS $|[l_i];v;\xi\rangle=
 S_V(\xi)^{\dagger}|[l_i];v; \rangle$ as trial functions.

Both ways above essentially exploit the finite-dimensionality of the $sl(2)$
adjoint (vector) representation (cf. Eq. (2.1a)) and well-known (due to
Eq. (2.1)) explicit expansions of the $SL(2)$ GCS in orthonormalized basis
states. However, for polynomial Lie algebras $sl_{pd}(2)$ the situation is
more complicated since their adjoint representations defined by repeated
commutations of arbitrary elements are infinite-dimensional as it follows
from Eq. (1.2). Furhtermore, GCS exponential operators  $S_V(\xi)=
\exp(\xi V_+-\xi^* V_-)$ have not explicit expressions for matrix elements
in orthonormalized bases (1.6) as these exponentials are not elements of
Lie groups but only correspond to quasigroups (pseudogroups) /10/ which have
no simple analogs of the "disentangling theorem" (2.1) providing expansions
of operators $S_V(\xi)$ in finite products of one-parameter subgroups /10,
11/. Therefore, in this case a direct generalization of results (2.3) is
impossible.

Nevertheless, taking into account Eqs. (1.2), (1.8) one may apply
the diagonalizing scheme (2.2) using repres¥ntations of diagonalizing
operators $S(\xi)$ by power series
%%%
$$S(\xi)=\sum_{f=-\infty}^{\infty}
V_+^f S_f(V_0;\xi), \quad V_+^{-k}\equiv V_-^k([\psi_n(V_0)]^{(k)})^{-1},
\; k>0     \eqno (2.5)$$
 %%%%%
with undetermined (unlike those for the $sl(2)$ algebra - cf. (2.1)
and (2.3)) coefficients $S_f(V_0;\xi)$ (which, when being known, provide
possibilities of explicit calculations of any physical quantities with the
help of Eqs. (1.2), (1.6)). For diagonalizing operators $S(\xi)=S_V(\xi)=
\exp(\xi V_+-\xi^* V_-)$ (if they exist) these coefficients may be taken
in the form $S_f(V_0;\xi)= \exp(i f\theta)\sigma_f(V_0;r),\;\xi=r
\exp(i\theta),$ and satisfy the equations
%%%%%
$$ \frac{\partial \sigma_f(V_0;r) }{\partial r} -\sigma_{f-1}(V_0;r)+
\psi_n(V_0+f)\sigma_{f+1}(V_0;r)=\delta(r) \delta_{f,0}, \quad  f=0,1,...
   \eqno (2.6)$$
%%%%%%
whose solutions may be represented by power series in $r$ (via direct
expansions of exponents $S_V(\xi)$) or obtained in an integral form with
the help of the "$sl(2)$ dressing" procedure /4/.
In general cases these coefficients satisfy the equations
%%%%%
$$ \sum_{f=-\infty}^{\infty} [\psi(V_0)]^{(f)} S_{k+f}(V_0-f;\xi)
S_f^*(V_0-f;\xi) =\delta_{k,0}
\eqno (2.7)$$
%%%%%
following from the unitarity conditions $SS^{\dagger} =S^{\dagger}S=I$.

Then, substituting Eq. (2.5) in the scheme (2.2) one  gets  after some
algebra nonlinear analogs of Eqs. (2.2)
%%%%%
$$a)\,\tilde{H}(\xi)=S_V(\xi)H S_V(\xi)^{\dagger}= C +
\sum_{f=-\infty}^{\infty}V_+^f \tilde{h}_f(V_0;\xi),\quad V_+^{-k}
\equiv V_-^k([\psi(V_0)]^{(k)})^{-1}, \; k>0,$$
$$ \tilde{h}_f(V_0;\xi)= \sum_{k=-\infty}^{\infty}[\psi(V_0)]^{(k-f)}
S_k(V_0+f-k;\xi) [a(V_0+f-k) S_{k-f}^*(V_0+f-k;\xi) +$$
$$g\psi(V_0+f-k)S_{k+1-f}^*(V_0+f-k-1;\xi)+ g^*S_{k-1-f}^*(V_0+f-k+1;\xi)],$$
$$ \tilde{h}_{-f}(V_0;\xi)=\tilde{h}^*_f(V_0-f;\xi)[\psi(V_0)]^{(f)},
\quad[\psi(V_0)]^{(-f)}\equiv([\psi(V_0+f)]^{(f)})^{-1},\quad f>0,
\eqno (2.8a)$$
$$b)\,\tilde{H}(\xi_0)=S(\xi_0)H S(\xi_0)^{\dagger}= C +
 \tilde{h}_0(V_0;\xi_0),\; E([l_i];v;\xi)= C+\langle[l_i];v|
 \tilde{h}_0(V_0;\xi)|[l_i];v\rangle         \eqno (2.8b)$$
%%%%%%%%%%
expressed in terms of the coefficients $S_f(V_0;\xi)$ (hereafter the
subscript $n$ in $\psi_n(V_0)$ will be omitted for the sake of the notation
simplicity). As is seen from Eq. (2.8b) the diagonalized Hamiltonian
$\tilde{H}(\xi_0)$  has (unlike (2.2b)) an essentially non-linear dependence
in  $V_0$ determined by coefficients $S_f(V_0;\xi_0)$ which satisfy
(additionally to Eqs. (2.7)) the operator recurrence relations following
from the condition $S(\xi_0)H=\tilde{H}(\xi_0)S(\xi_0)$),
%%%%%%
$$ S_f(V_0;\xi_0)[aV_0 - \tilde{h}_0(V_0+f;\xi_0)] +g S_{f-1}(V_0+1;\xi_0) +
g^* S_{f+1}(V_0-1;\xi_0)=0, \; f=0,\pm 1,\pm 2,..,$$
$$ \tilde{h}_0(V_0;\xi_0)=aV_0+ \sum_{n=-\infty}^{\infty}[\psi(V_0)]^{(n)}
S_n(V_0-n;\xi_0) [-n S_{n}^*(V_0-n;\xi_0) +$$
$$g\psi(V_0-n)S_{n+1}^*(V_0-n-1;\xi_0) + g^*S_{n-1}^*(V_0-n+1;\xi_0)]
                    \eqno (2.9a)$$
%%%%%%%%%%%%%%%%%%%%%%%%
or the operator equations
%%%%%%%%%%%%%%%%%%%%%%
$$ 0 =\sum_{n=-\infty}^{\infty}[\psi(V_0)]^{(n-f)}
S_n(V_0+f-n;\xi_0) [a(V_0+f-n) S_{n-f}^*(V_0+f-n;\xi_0) +$$
$$g\psi(V_0+f-n)S_{n+1-f}^*(V_0+f-n-1;\xi_0) + g^*S_{n-1-f}^*(V_0+f-n+1;
\xi_0)],\; f=\pm 1,\pm 2,..
                    \eqno (2.9b)$$
%%%%%%%%%%%%%%%%%%%%%%%%
resulting from the condition $\tilde{h}_f(V_0;\xi_0)=0,\; f=\pm 1,\pm 2,..$
(a direct generalization of the condition $A_+(a,g;\xi)=0$ in (2.2)). Note
that Eqs. (2.9), in general, determine both a suitable functional form of
$S_f(V_0;\xi)$ and a value $\xi_0$ of the parameter $\xi$ diagonalizing
the Hamiltonian $\tilde{H}(\xi)$.

So, the formalism of the $sl_{pd}(2)$ algebras enabled to to get a general
operator scheme  of diagonalizing the Hamiltonians (1.1) with the help
of solving the (infinite) set  of algebraic operator equations (2.7)-(2.9).
Evidently, without using some specifications of diagonalizing operators
$S(\xi)$ the task of solving these equations is equivalent to that for
finding amplitudes $Q_v(E_f)=\langle [l_i]; v|S^{\dagger}|[l_i]; f \rangle=
[(\psi_n(l_0+f))^{(f)}/(\psi_n(l_0+v))^{(v)}]^{1/2} S_{f-v}^*(l_0+v;\xi_0)$
as Eqs. (2.9b) resemble those for $Q_v(E_f)$. Note that in the case of the
compact $su_{pd}(2)$ algebra, having finite-dimensional (with dimensions
equal to $d([l_i])$) irreducible subspaces $L([l_i])$, it is possible to
simplify the task restricting oneself by the consideration of Eqs.
(2.7)-(2.9) on each $L([l_i])$ {\bf independently}. Then, due to the
relation $(V_{\pm})^{d([l_i])+1}|_{L([l_i])} =0$  all
series in Eqs. (2.5), (2.7)-(2.9) are terminating. Specifically, wave
eigenfunctions $|E_f \rangle=S^{\dagger}|[l_i];f \rangle$ may be represented
by polynomials $|E_f \rangle= A(V_0)\prod_j (V_+ -\Lambda _j(V_0))$
and the energy functionals $E([l_i]; f)_S\equiv \langle [l_i];f |S H
S^{\dagger}|[l_i]; f \rangle$ are written down in the form of a sum of
$d([l_i])$ spectral functions as it is prescribed for such classes of models
by the algebraic Bethe ansatz /12/. (In essence, we  obtain in such a manner
a new formulation of this ansatz for a wide class of models in terms of the
$su_{pd}(2)$ algebras which is simpler and more efficient (cf. /3/) in
comparison with its initial version /12/.)

However, even such simplifications enable us to get simple closed expressions
only for little dimensions $d([l_i])$ of the $su_{pd}(2)$ irreducible
subspaces $L([l_i])$. At the same time many physical quantum states of models
(1.1), e.g., such as coherent and squeezed states in  models (1.3), have
non-zero projections on all subspaces $L([l_i])$. Therefore, for physical
applications it is necessary to get some closed expressions like Eqs. (2.3)
for energy eigenvalues and wave eigenfunctions which would describe main
features of model dynamics with a good accuracy. One exampl¥ of such
analytical approximations was obtained in /1-3/ by mapping (with the help of
the change $V_{\alpha}\rightarrow Y_{\alpha}$) Hamiltonians (1.1) in
Hamiltonians $H_{sl(2)}$ which are linear in $sl(2)$ generators $Y_{\alpha}$
and have on each fixed subspace $L([l_i])$ equidistant energy spectra given
by formulas like Eq. (2.3a) (but with modified constants $a,g$). However,
this (quasi)equidistant approximation is suitable for little or very big
dimensions $d([l_i])$ and does not enable to display many peculiarities (e.g.,
availability and a fine structure of collapses and revivals of the Rabi
oscillations) of models (1.1). Therefore, below we describe an alternative
approximation applying the variational scheme (2.4) with $SL(2)$ GCS as
trial functions to Hamiltonians (1.1) expressed with the help of Eqs. (1.8)
as functions of $sl(2)$ generators $Y_{\alpha}$.

\section{A variational scheme of determining energy spectra with the
help of $SL(2)$-coherent states}

Hamiltonians (1.1) re-written in terms of $Y_{\alpha}$ have the form
%%%
$$ H = aY_0 +g Y_+\sqrt{\frac{\psi_n(V_0+1)}{(j\mp Y_0)(\pm j+1+Y_0)}} +
g^* \sqrt{\frac{\psi_n(V_0+1)}{(j\mp Y_0)(\pm j+1+Y_0)}}Y_- +
C+a(\pm j+l_0)      \eqno (1.1') $$
%%%%%
which is essentially  non-linear in $sl(2)$ generators $Y_{\alpha}$.
Therefore, in general, it is unlikely to diagonalize them with the help of
operators $S_Y(\xi)=\exp(\xi Y_+-\xi^* Y_-)$.
However, it is natural to apply associated with these operators $SL(2)$ GCS
%%%%
$$|[l_i];v;\xi\rangle= S_Y(\xi)^{\dagger}|[l_i];v; \rangle= \exp(-\xi Y_+
+ \xi^*Y_-)|[l_i];v; \rangle,      \eqno (3.1)$$
%%%%%%%
as trial functions  in the variational scheme (2.4) that results in
non-linear analogs of Eq. (2.3a) for approximate energy eigenvalues. Such
an approximation may be called as a "smooth" $sl(2)$ approximation
since it, in fact, corresponds to picking out a "smooth" (due to analytical
nature of $SL(2)$ group elements) $sl(2)$ factor $\exp(\xi Y_+-
\xi^* Y_-)$ in the exact diagonalizing operators $S(\tilde{\xi})$.

Specifically, application of this procedure to Hamiltonians with the
$su_{pd}(2)$ dynamic symmetry yields after some algebra the following
expressions
%%%%%
$$ E([l_i];v;\xi_0) =C+a(l_0+j)+a(-j+v)\cos 2r-2|g|(\cos^2 r)^{2(j-v)}
\frac{(2j-v)!}{v!}\sum_{f\geq 0} E_v^{\psi}(l_0,j;f),$$
$$E^{\psi}_v(l_0,j;f)=\frac{(tg r)^{2(f-v)+1}(f+1)!}
{(f-v)!(f+1-v)!(2j-f-1)!}\sqrt{\frac{\psi(l_0+1+f)}{(2j-f)(f+1)}}\times$$
$$F(-v,-v+2j+1;f-v+1; \sin^2 r)F(-v,-v+2j+1;f-v+2; \sin^2 r)  \eqno (3.2)$$
%%%%%%
for energy eigenvalues  $E([l_i];v;\xi_0=re^{i\theta})$ where
$e^{i\theta}= g/|g|$ due to the second condition (2.4) and diagonalizing
values of the parameter $r$ are determined from solving the equations
%%%%
$$0 = \sum_{f\geq 0}\frac{\alpha^{2f}}{(2j-1-f)!f!} \{\frac{a \alpha}{|g|}-
[4\alpha^2 j -(1+\alpha^2)(2f+1)]\sqrt{\frac{\psi(l_0+1+f)}{(2j-f)(f+1)}}\},
\;\alpha =-tg r \eqno (3.3)$$
resulting from the first condition (2.4).

As is seen from Eq. (3.2), spectral functions
$E^{\psi}_v(l_0,j;f)$ are non-linear in the discrete variable $v$
labeling energy levels that provides a non-equdistant character of energy
spectra within  fixed subspaces $L([l_i])$ at $d([l_i])>3$; besides,
due to the availability of square roots in expressions for these functions
different eigenfrequencies $\omega_v\equiv E_v/\hbar$ are incommensurable:
$m\omega_{v_1}\neq n\omega_{v_2}$ that is an indicator of an origin
of collapses and revivals of the Rabi oscillations /5/ as well as of
prechaotic dynamics. We also note that
Eqs. (3.2)-(3.3) give exact results at little dimensions $d([l_i])$.

\section{Conclusion}

So, we have obtained a general operator scheme for diagonalizing Hamiltonians
(1.1) and a smooth approximation for solutions of its defining equations with
the help of the mapping (1.8) and the variational scheme (2.4) using the
$SL(2)$ GCS as trial functions. This approximation may be used as an initial
one in iterative schemes of solving Eqs. (2.9) (re-written in "the $sl(2)$
languauge") which are similiar to those developed to examine non-linear
problems of classical mechanics and optics /11/. Further investigations
along this line may be also related to a search of suitable specifications
of the operators $S(\xi)$ (besides the form $S(\xi)=S_V(\xi)$) reducing
solutions of Eqs. (2.9) to determining a value $\xi_0$  providing an exact
or a sufficiently accurate approximation for diagonalization of the
Hamiltonian (1.1) in scheme (2.3a). Another way to develop the results above
concerns some simplifications of the formulas (3.2) via using different
properties, including integral representations, of the hypergeometric
functions $F(a,b;c;x)$. For the case of $\psi_3(x)$ it is also of interest to
compare results of such approximations with exact calculations obtained
by considering exactly solvable cases of the Riccati equations yielded by
the $sl_{pd}(2)$ representation (1.7b).

Finally, general ideas of the analysis above may be extended to solve
evolution problems. Specifically, a version of general operator formalism
for these tasks was formulated in /4/, and a version of obtaining a
variational dynamics in the mean-field approximation can be found following
the approach of the paper /8/. Namely, looking for the evolution operator
$U_H(t)$ (with $H$ given by Eq. (1.1')) in the form $U_H(t)=
\exp(-z(t) Y_++ z(t)^*Y_-)$ and using the $sl(2)$ GCS $|z(t)\rangle=
\exp(-z(t) Y_++ z(t)^*Y_-)|\psi_0\rangle$ as trial functions in the
time-dependent Hartree-Fock varitional scheme with the Lagrangian ${\cal L}
=\langle z(t)|(i\partial/\partial t - H)|z(t)\rangle$ one gets the $sl(2)$
Euler-Lagrange equations
%%%%%
$$ \dot q =\frac{\partial {\cal H}}{\partial p}, \qquad \dot p =
-\frac{\partial {\cal H}}{\partial q} \eqno (4.1)$$
%%%%%%%%
for "motion" of the $sl(2)$ GCS parameters; here ${\cal H}=
\langle z(t)|H|z(t)\rangle$ and $p=j\cos \theta, q= \phi,
z = \tan (\theta/2) \exp (-i\phi)$ for $su(2)$ and $p=j\cosh \theta, q= \phi,
z = \tanh (\theta/2) \exp (-i\phi)$ for $su(2)$.

\section{Acknowledgement}

The author thanks A.V. Masalov, S.M. Chumakov and M.A. Mukhtarov for useful
discussions and G.S. Pogosyan for interest in the work. A partial financial
support of participation in the Conference from its Organizing Committee is
acknowledged.

\end{document}